\documentclass[12pt]{article}
\newlength{\dinwidth}
\newlength{\dinmargin}
\begin{document}
\noindent Submitted to Journal of Physics G \\

\begin{center}
\begin{LARGE}
   The Future of Lepton-Nucleon Scattering  \\
\end{LARGE}
\begin{Large}
a summary of the Durham workshop, December 2001\\
\end{Large}
\vspace{1cm} T.~Greenshaw (Liverpool) and M.~Klein (DESY-Zeuthen)

24$^{\rm th}$ April 2002

\end{center}
\vspace{0.5cm} A summary is given of the recent Durham workshop on the future of
lepton-nucleon scattering. Discussions at the workshop centred on the need to ensure
that an international scientific and technical programme is developed with the goal
of further exploring the structure of the nucleon. Questions of particular interest
include the investigation of nucleon structure and spin at extremely small Bjorken
$x$. The role of the Electron Ion Collider in this programme was discussed, as was
the necessity of ensuring that HERA is fully exploited. \vspace{0.5cm}
\section{The Situation - Open Questions after HERA2}
HERA is currently scheduled to run until the end of 2006, the aim being to collect
polarised $\overrightarrow{e}p$ scattering data corresponding to an integrated
luminosity of 1~fb$^{-1}$ at the highest possible energy and to obtain some data at
lower energy. This programme will allow exploration of proton structure at the
smallest possible dimensions and will also produce large amounts of precision data
at low Bjorken $x$. These data may result in the discovery of new effects at low $x$
or at  high momentum transfers. In the latter case, the comparison of precision
measurements with predictions from the QCD evolution equations, well understood in
this kinematic regime, leads to sensitivity to the effects of large extra dimensions
or contact interactions in the multi-TeV range. The production of new heavy
particles, such as the squarks of $R$-parity violating supersymmetric theories,
leptoquarks or excited fermions is also possible up to masses of about $320$ GeV and
the effects of these particles can be detected for masses significantly above this.
The observation of novel effects with the envisaged five-fold increase in $e^+p$ and
thirty-fold increase in $e^-p$ integrated luminosity would of course make further
investigations beyond the year 2006 mandatory. Even if no new effects are
discovered, the current HERA programme will leave a number of fundamental questions
unanswered. Issues of prime importance are:
\begin{itemize}
\item{What is the partonic structure of the neutron at low $x$,
and at large $Q^2$ and $x$?}
\item{What is the origin of confinement in long distance (small $x$) and
short distance (large $x$) processes?}
\item{How does the total deep inelastic scattering cross-section for lepton-proton
and lepton-nucleus interactions change in the high density regime where cross
sections may saturate and how are these changes reflected in the hadronic final
state?}
\item{How do the partons conspire to ensure that the spin of the nucleon is 1/2?}
\end{itemize}
Understanding these problems will require data from lepton-deuteron and lepton-heavy
ion scattering, as well as a programme of high energy and high luminosity colliding
beam and fixed target experiments involving polarised leptons, polarised protons and
polarised deuterons. In addition to addressing the above questions, this programme
will be vital to our understanding of the astrophysical significance of extremely
high-energy neutrino scattering, as well as being necessary to ensure full
exploitation of the physics potential of the LHC and of heavy ion collisions.

The Durham meeting brought together about 40 physicists, including the
Spokespersons, Physics and Technical Coordinators of the HERA experiments H1, ZEUS,
and HERMES, of COMPASS at CERN, leaders of the Electron-Ion Collider (EIC) Community
and leading machine and theory experts. The programme of the workshop (available at
http://hep.ph.liv.ac.uk/$\sim$green/HERAfuture) comprised sessions devoted to the
current HERA programme (HERA2), QCD, polarised $\ell N$ physics, detectors,
deuterons and heavier nuclei and machine developments at HERA and the EIC. It ended
with an extended discussion about the plans, options and prospects for future
deep-inelastic scattering experiments prior to the operation of TESLA.
\section{Physics Subjects - Nuclei and Polarisation}
Much of the physics of nuclei and polarisation has been studied at previous
workshops, and frequent reference was made to the HERA $eA$ workshops~\cite{eA}, to
the HERA spin workshops~\cite{spin}, to the EIC white book~\cite{eic}, to the THERA
book~\cite{thera}, to the TESLA-N~\cite{tesla-n} and to the ELFE~\cite{elfe}
proposals.

As mentioned above, after completion of the currently approved HERA programme, there
will be a continued interest in precision measurements in the low $x$ region. This
concerns in particular the measurement of jets very close to the proton beam
direction, crucial to the understanding of the emission of gluons at low $x$, and
the kinematic region in which $Q^2 ~\sim 1$~GeV$^2$, where the energy dependence of
the $\gamma^* p$ cross-section becomes hadron-hadron like. Precision inclusive and
exclusive cross-section measurements in this kinematic region may yield insight into
saturation physics and the confinement problem. This programme will require that the
instrumentation in the forward and backward regions very close to the beam pipe at
HERA be upgraded, that efficient proton tagging is possible and that the beam
divergence is small.

Electron-deuteron scattering appears to be the natural next step at HERA following
the $ep$ programme.  Deuterons are a source of quasi-free neutrons, measurements of
which are perturbed slightly in the region $x < 0.1$ by nuclear shadowing effects.
These latter are related to the diffractive parton densities and hence the effects
of shadowing on lepton-deuteron cross-sections can be determined to an accuracy of
better than 1-2\%. The quark distribution asymmetry, $(u + \overline{u}) - (d +
\overline{d})$, can thus be accurately measured from the difference $F_2^p - F_2^n$.
Deuteron data are also essential for determining the individual flavour
decomposition of the nucleon parton distributions at large $x$, allowing measurement
of quantities such as $s - c$ or the $d/u$ ratio, as well as for determining charged
current structure functions and for precision tests of $Q^2$ evolution in
perturbative QCD.

The rise of the proton structure function $F_2$ towards low $x$ in the
deep-inelastic scattering region is due to the high sea quark density in the proton.
This is related to the gluon distribution $xg \propto \partial F_2 / \partial \ln
Q^2$, which also rises as $x^{-\lambda}$ towards low $x$, with $\lambda \approx 0.1
\dots 0.4$ for $Q^2 \approx 1 \dots 100$ GeV$^2$. Since $xg$ is expected to increase
with atomic number $A$ as $A^{1/3}$ (modulo shadowing effects), electron-nucleus
scattering allows an equivalent Bjorken $x = x_N /(A^{1/3})^{1/\lambda}$ to be
accessed: such collisions can thus be used to investigate a kinematic regime which
in $ep$ scattering would require a considerable increase in the centre-of-mass
energy. Diffractive processes may constitute up to 50\% of the inclusive
lepton-nucleus cross section, the maximum value allowed by unitarity considerations.
Such an observation would represent an unambiguous signal for a new regime in
deep-inelastic scattering. Since the contribution of small-size configurations to
diffraction goes as the square of $xg$, non-linear effects which damp the growth of
the total cross-section may first be seen in diffraction. The nucleus is widely
considered to be an ideal laboratory for deconfinement and high density QCD
analyses.

It is envisaged that spin physics with fixed polarised targets will be continued at
DESY beyond 2006 with a high luminosity measurement programme. With modest
improvements to the HERMES apparatus, high resolution experiments with increased
luminosity will become possible. Such measurements will be devoted to a detailed
study of exclusive reactions, in particular deeply virtual Compton scattering, and
will lead to the first precise experimental information on generalised parton
distributions.

Collisions of beams of polarised electrons and polarised nucleons will make possible
the first investigations of spin phenomena at high $Q^2$ and at low $x$. These will
be enhanced by the comprehensive reconstruction of the final state possible in
colliding beam experiments. The behaviour of the spin structure function
$g_1(x,Q^2)$ at low $x$ is unknown, but it is expected to change even more
dramatically than $F_2$.  The $Q^2$ dependence of $g_1$ determines the gluon spin
distribution $\Delta G$, one of the components of the spin of the proton.
Semi-inclusive measurements can be used to explore the flavour structure of spin:
dijets give access to $\Delta G$ and deeply virtual Compton scattering to the
generalised parton distributions. Photoproduction processes give insight into the
polarised gluonic structure of the photon and diffractive polarised scattering into
pomeron exchange. Beyond the fixed target $\overrightarrow{e}\overrightarrow{p}$
domain, there is a vast unexplored kinematic region  and polarised colliding beam
experiments may radically change our view on the spin structure of the nucleon.
\section{Detector Aspects}
The status and upgrade plans for H1, ZEUS and HERMES at DESY, and COMPASS at CERN
were summarised by the technical coordinators of these experiments. It was pointed
out that in 2006 the HERA collider detectors will be 15 years old. Continued
operation will require some detector and electronics upgrades. For example, whereas
the H1 liquid argon calorimeter has proved to be extremely stable, in all
probability an upgrade to the central tracking facilities will be necessary,
including replacement and possibly extension of the silicon detectors. More accurate
judgements on the long-term prospects for operation of the H1 and ZEUS detectors
will be possible following the first years of high luminosity running at HERA2.

Currently, the HERA interaction regions are optimised for high luminosity running,
with focussing magnets placed close to the interaction region. This sets a limit on
the minimum measurable four-momentum transfer squared of about 4~GeV$^2$.  These
high luminosities are essential for the measurement of small spin-induced
asymmetries and also for high statistics $eD$ measurements. At low $Q^2$, event
rates will be high and substantial upgrades of the front-end electronics may be
required.

For accessing the region around $Q^2=1$ GeV$^2$ and the smallest $x$ values in $eA$
scattering, it will be necessary to return to an arrangement which leaves room for
new or upgraded detectors close to the beam pipe.  An essential requirement for
diffractive measurements and for deuteron spectator tagging is large acceptance
forward nucleon tagging.

It is expected that the HERMES detector, perhaps with some upgrades to the data
acquisition system and silicon detectors, will be able to run beyond 2006.  The
COMPASS experiment uses a variety of modern detector technologies which may be of
use for future HERA3 detectors.

\section{Accelerators - HERA and the EIC}
Following the recent upgrade, HERA is scheduled to run with a luminosity of
$7~\times~10^{31}$ cm$^{-2}$s$^{-1}$ until the end of 2006. The maximum attainable
luminosity is estimated to be $13 \times 10^{31}$ cm$^{-2}$s$^{-1}$. This
corresponds to an annual integrated luminosity of about 500~pb$^{-1}$, making the
measurement of small asymmetries at low $x$ in polarised $\overrightarrow{e}
\overrightarrow{p}$ scattering statistically feasible if intense polarised proton
sources can be constructed. Precise measurements require a high degree of
polarisation which must be transferred through the accelerator chain, necessitating
the use of Siberian Snakes. Polarimeters allowing accurate polarisation measurements
are also essential. The small anomalous magnetic moment of the deuteron may make
realisation of polarised $\overrightarrow{e} \overrightarrow{D}$ collisions easier
than the $\overrightarrow{e} \overrightarrow{p}$ case using a scheme which relies on
resonant driving of the spin using horizontal RF fields.

Pilot $eA$ scattering studies can be performed using deuterium, oxygen and calcium.
Light ions can be accelerated in HERA with moderate modifications to the accelerator
chain. Electron cooling becomes necessary if heavier nuclei are to be accelerated to
counter intra-beam scattering (IBS) effects. The luminosity is then expected to
scale as $L_A \simeq L_p /A$. For deuterons, the IBS time exceeds two hours and a
luminosity of $3.5 \times 10^{31}$ cm$^{-2}$s$^{-1}$ may be achieved without
cooling. Thus, high luminosity $eD$ running can be realised immediately after
completion of the HERA2 programme and requires no major modifications to HERA beyond
the necessary changes to the source.

The EIC, formal proposals for which are expected to be presented in about 2005, will
be an intense polarised electron-ion collider using electron beam cooled ions in
RHIC (of up to $E_p = 250~$GeV proton and $E_{Au} =100~$GeV/A gold energy) colliding
with electrons from a ring or linear accelerator of about 10~GeV maximum energy.
Thus the EIC has 10 times less centre-of-mass energy than HERA, but due to its high
luminosity it greatly extends the range of current polarised fixed target
experiments, and with heavy nuclei allows access to high parton densities. The
ring-ring accelerator has an estimated luminosity of $25~(0.7) \times 10^{31}$
cm$^{-2}$s$^{-1}$ for protons (gold), which is comparable to HERA after run 2. With
an energy recovery linac, it is expected that a still higher luminosity of $100
~(1.0) \times 10^{31}$ cm$^{-2}$s$^{-1}$ for protons (gold) may be obtained. The
linac, while accelerating $e^-$ only, has the further advantage of providing high
polarisation and avoiding synchrotron radiation background.
\section{Conclusions and Outlook}
If answers are to be found to the questions posed in the introduction, three
experimental programmes must be pursued:
\begin{itemize}
\item{Electron-deuteron scattering experiments, with both high luminosity
(more than 100~pb$^{-1}$) and detectors that have dedicated low $x$ and forward
nucleon tagging capability.}
\item{Extended measurements at low $x$: studies of the problems of confinement require
precision diffractive and
non-diffractive data at $Q^2$ near to 1~GeV$^2$; high density gluon effects leading
to saturation may be studied in electron-nucleus scattering.}
\item{High luminosity polarised $\overrightarrow{e} \overrightarrow{p}$ and
$\overrightarrow{e} \overrightarrow{D}$ scattering: spin physics requires further
operation of fixed target experiments and the investigation of the unknown world of
low $x$ and high $Q^2$ deep-inelastic spin phenomena using collider detectors.}
\end{itemize}
Work in the coming years will be directed towards producing proposals for the
operation of HERA beyond 2006 and for the EIC which may  start operation in 2012,
after completion of the polarised proton phase of RHIC. A programme will be devised
which produces maximum physics return from HERA and the EIC, with their
complementary reach in energy, luminosity and nuclear mass.  An extension by another
order of magnitude in energy in $ep$ scattering can be achieved with THERA, the
future $ep$ collider operating at TeV energies.

\section{Acknowledgements}
The authors would like to thank the DESY Directorate and the Institute for Particle
Physics Phenomenology for their support and Linda Wilkinson and the other members of
the organising committee for helping to arrange what proved to be a most interesting
and productive meeting, namely Jochen Bartels, Alan Martin, Martin McDermott, Klaus
Rith, Jim Whitmore, Ferdinand Willeke and Giulia Zanderighi. This summary has
benefited from the suggestions of Allen Caldwell, Eckhard Elsen, Robert Klanner,
Alan Martin, Martin McDermott, Dirk Ryckbosch, Peter Schleper, Mark Strikman and
Ferdinand Willeke. We would also like to thank the representatives of the theory
community and the COMPASS, HERMES, H1 and ZEUS Collaborations who contributed to the
workshop.

\end{document}